\documentclass[12pt]{iopart}
\usepackage[all]{xy}
\usepackage{graphicx}
\usepackage{subfigure}
\usepackage{color}
\usepackage{iopams}  
\begin{document}

\title{The effect of reactions on the formation and readout of the gradient of Bicoid}

\author{Emiliano Perez Ipi\~na$^{1,2}$ \footnote{To whom correspondence should be addressed (emperipi@df.uba.ar)} and Silvina Ponce Dawson$^1$}

\address{$^1$Departamento de
  F\'\i sica, FCEN-UBA, and IFIBA, CONICET,
    Ciudad Universitaria,
  Pabell\'on I, (1428) Buenos Aires, Argentina\\
  $^2$Laboratoire J. A. Dieudonn\'e Universit\'e de Nice Sophia Antipolis, UMR 7351 CNRS, Parc Valrose, F-06108 Nice Cedex 02, France}

\begin{abstract}
During early development, the establishment of gradients of
transcriptional factors determines the patterning of cell fates. The
case of Bicoid (Bcd) in {\it Drosophila melanogaster} embryos is well
documented and studied. There are still controversies as to whether
{\it SDD} models in which Bcd is {\it Synthesized} at one end, then
{\it Diffuses} and is {\it Degraded} can explain the gradient
formation within the timescale observed experimentally. The Bcd
gradient is observed in embryos that express a Bicoid-eGFP fusion
protein (Bcd-GFP) which cannot differentiate if Bcd is freely
diffusing or bound to immobile sites.  In this work we analyze an {\it
  SDID} model that includes the {\it Interaction} of Bcd with binding
sites.  Using previously determined biophysical parameters we find
that this model can explain the gradient formation within the
experimentally observed time.  Analyzing the differences between the
free and bound Bcd distributions we observe that the latter spans over
a longer lengthscale. We conclude that deriving the lengthscale from the
distribution of Bcd-GFP can lead to an overestimation of the gradient
lengthscale and of the degree of cooperativity that explains the distribution
of the protein Hunchback whose production is regulated by Bcd.
\end{abstract}

\vspace{2pc}
\maketitle

\section{Introduction} 
During the early development of embryos, cell differentiation is
carried out by the transcriptional regulation of gene expression. The
establishment of gradients of morphogens determines the patterns of
cell fates. Cells get the information on their relative spatial
location by ``reading'' the local concentration of these
morphogens~\cite{Wolpert1969,Crick1970}. \textit{Drosophila
  melanogaster} embryos constitute a model system in which this
subject has been studied with great detail. In this system, bicoid
(\textit{bcd}), a maternal effect gene whose mRNA is localized at the
anterior end of the embryo~\cite{st_johnston1989, Little2011}, plays a
key role. The protein, Bcd, one of the principal morphogenes in these
embryos, is responsible, in conjunction with other factors, of the
anterior-posterior (AP) axial patterning. \textit{bcd} is translated
into Bcd mainly at the anterior pole of the embryo~\cite{Little2011},
forming a concentration gradient along the anterior-posterior
axis. Bcd is a transcription factor for over 20 target genes involved
in the development. In particular, its role in the regulation of
Hunchback is fundamental during the early embryogenesis of
\textit{Drosophila}.  After fertilization, the cell undergoes several
nuclear division cycles (n.c) without cytokinesis. After n.c. 7,
nuclei move to the surface forming a syncytial blastoderm and $\sim 4$
hours after fertilization, just before cytokinesis begins (close to
n.c. 14) there are approximately 6000 nuclei on the surface. Recently,
live imaging using Bcd-GFP allowed the observation of the
spatio-temporal distribution of Bcd during these early stages of the
embryo development~\cite{Gregor:Cell2007a}. From these observations it
was determined that the Bcd concentration gradient is established
within the first 10 n.c., {\it i.e.}, 90 minutes after egg
deposition. Then, between n.c 10 and 14 the gradient remains almost
unchanged and thereafter begins to decrease.
The mechanisms by which the Bcd gradient is established so early are
still not completely determined~\cite{Grimm2010}.

One of the simplest and most widely used models to explain the
formation of the Bcd gradient is the {\it Synthesis}, {\it Diffusion},
{\it Degradation} (SDD)
model~\cite{Driever1988,Nusslein1989,Strhulet1989}. It assumes that
Bcd is synthesized at a constant rate, $\zeta$, at the anterior
end, then diffuses along the antero-posterior axis ($z$) of the embryo
with diffusion coefficient, $D$, while it is being uniformly degraded
with rate, $\alpha$. Assuming, for simplicity, a cylindrical embryo
of transverse area, $A$, and total length, $L$, the dynamic equation of the Bcd concentration,
$[Bcd]$, in the SDD model can then be written as:
\begin{equation}
\frac{\partial{[Bcd]}}{\partial t} = D\nabla^2 [Bcd]  -{\alpha}[Bcd]
  ,\label{eq:ec_sdd}
\end{equation}
with boundary conditions:
\begin{eqnarray}
  -D\frac{\partial{[Bcd]}}{\partial z} A &=& \zeta, \quad
  z=0\nonumber\\ \frac{\partial{[Bcd]}}{\partial z}&=& 0, \quad z=L
  .\label{eq:ec_sdd_bc}
\end{eqnarray}
For long enough $L$ the solution of (\ref{eq:ec_sdd})--(\ref{eq:ec_sdd_bc})
coincides with that of:
\begin{equation}
\frac{\partial{[Bcd]}}{\partial t} = D\nabla^2 [Bcd]  -{\alpha}[Bcd] +\frac{2\zeta}{A}\delta(z)
  ,\quad\quad \frac{\partial{[Bcd]}}{\partial z}=0\, {\rm for}\, z\rightarrow \pm\infty\label{eq:ec_sdd_point_source}
\end{equation}
in the region $z\ge 0$. The stationary solution of
(\ref{eq:ec_sdd})--(\ref{eq:ec_sdd_bc}) is:
\begin{equation}
\left[Bcd\right](z) = [Bcd]_o exp(-z/z_o),
\label{eq:SDD}
\end{equation}
where $[Bcd]_o=\zeta/(A\sqrt{\alpha D})$ and
\begin{equation}
z_o = \sqrt{D/\alpha}, \label{eq:long_car_sdd}
\end{equation}
is the characteristic length-scale of the
gradient~\cite{Houchmandzadeh2002}.  Observations of the Bcd
distribution have led to the estimate $z_o\sim 100\mu m$.  Within the
framework of the SDD model, the time it takes for Bcd to diffuse over
this distance is of the order of $100\mu m^2/D$. Thus, for the
gradient to be established within 90 min it is necessary that $D\ge
10^4\mu m^2/(90min)\sim 2\mu m^2/s$~\cite{Gregor:Cell2007a}.  The
first quantification of the Bcd diffusion coefficient was obtained
using Fluorescence Recovery After Photobleaching (FRAP) during
mitosis~\cite{Gregor:Cell2007a}. The estimated value was $D_{FRAP}
\sim 0.3 \mu m^2/s$ which was an order of magnitude too small to
explain the establishment of the Bcd gradient within the SDD model
during the experimentally observed times. In~\cite{Bergmann2007} it
was argued that after nc 14 the gradient does not reach the steady
state solution, so that (\ref{eq:SDD}) is not valid to estimate
the diffusion rate. Other alternatives to the SDD model have also been
proposed that involve an active~\cite{Gregor:Cell2007b} or
advective~\cite{Hecht2009} transport of the Bcd protein. Another model
stated that the stability of the gradient between n.c. 10 and 14 could
be explained in terms of an underlying mRNA
gradient~\cite{Spirov2009}. The distribution of mRNA was measured
in~\cite{Little2011} finding that it is bell-shaped with an 80\%
concentrated within the 20\% of the total embryo's length that lies
closest to the anterior pole. This seems to discard the possibility
that the Bcd gradient is a simple reflection of the way its mRNA is
distributed along the axis.  Regarding diffusion, the Bcd coefficient
was measured again more recently using Fluorescence Correlation
Spectroscopy (FCS) in the cytoplasm during
interphase~\cite{AbuArish:BJ2010} and in nuclei~\cite{Porcher2010}. The value
estimated in the former was $D_{FCS} \sim 7 \mu
m^2/s$~\cite{AbuArish:BJ2010}. In that work the diffusion coefficient
$D$ was also estimated using FRAP obtaining $D_{FRAP} \sim 1 \mu
m^2/s$. Although the FCS result apparently reconciles the observed
time it takes for the gradient to be formed with the SDD model, the
question arises as to what is the reason for the discrepancy between
$D_{FCS}$ and $D_{FRAP}$.  In~\cite{Sigaut2014} an explanation was
provided for this apparent discrepancy.

Based on previous studies on the transport of substances that diffuse
and react~\cite{Pando2006} and on the analysis of FCS and FRAP
experiments in such a case~\cite{Sigaut2010}, it was shown
in~\cite{Sigaut2014} that both the FRAP and FCS estimates of the Bcd
diffusion coefficient~\cite{Gregor:Cell2007a,Porcher2010} could be
correct if the interaction of Bcd with immobile or slowly moving
binding sites was taken into account. Namely, if Bcd is assumed to
diffuse with free coefficient, $D_f$, and react with more massive
binding sites, $S$, according to:
\begin{equation}
Bcd + S\,\,\raisebox{-2.5ex}{$\stackrel{\stackrel{\textstyle k_{on}}
{\textstyle{\longrightarrow }}}{\stackrel{\textstyle{\longleftarrow}}
{\textstyle{k_{off}}}}$} \,\, Bcd_b,
\label{eq:reaction_scheme}
\end{equation}
its net transport can be described in terms of ``effective'' diffusion
coefficients. As shown in~\cite{Pando2006} the effective coefficient
is different depending on whether one looks at the transport of a
single molecule in which case it is:
\begin{equation}
D_{sm} \equiv \frac{D_f + \frac{S}{K_d} D_S}{1+\frac{S}{K_d}}
\label{eq:Dsm}
\end{equation}
or at the dispersion of a collection of them in which case it is:
\begin{equation}
D_{coll} \equiv \frac{D_{f} + \frac{[S]^2}{K_d [S_T]} D_S}{1 + \frac{[S]^2}{K_d [S_T]}}.
\label{eq:Dcoll}
\end{equation}
In (\ref{eq:Dsm})--(\ref{eq:Dcoll}), $D_S$ is the free diffusion
coefficient of $S$ and of $Bcd_b$, $K_d\equiv k_{off}/k_{on}$ is the
dissociation constant of the reaction (\ref{eq:reaction_scheme}) and
$[S]$ and $[S_T]$ are the free and the total binding site
concentrations.  FRAP provides an estimate of
$D_{sm}$~\cite{Sprague2005} and FCS gives estimates of $D_{sm}$ and
$D_{coll}$~\cite{Sigaut2010}. Interpreting the results
of~\cite{Gregor:Cell2007a,Porcher2010} within this framework the work
of~\cite{Sigaut2014} showed that the FRAP and the FCS estimates of the
Bcd diffusion coefficient were compatible. Associating the estimated
coefficients to $D_{sm}$ or $D_{coll}$, depending on the experiment,
the concentrations and the dissociation constant were
determined~\cite{Sigaut2010}. In the current paper we use the same
framework with the parameters determined in~\cite{Sigaut2010} to
analyze how $[Bcd]$ changes with space and time along the embryo when
the {\it Interaction} with binding sites is considered as well as its
{\it Synthesis}, {\it Diffusion} and {\it Degradation}.  Thus, we
analyze the formation of the Bcd gradient within an {\it SDID model}.
It is likely that Bcd binds cooperatively to binding sites and not as
prescribed by~(\ref{eq:reaction_scheme}). Our {SDID} model should
then be interpreted as a toy model where to investigate how the
characteristic length and time scales of $[Bcd]$ are affected when the
interaction with binding sites is considered. In spite of its
simplicity, its predictions can be contrasted with the
observations. Furthermore, it helps pinpoint the main drawbacks of
interpreting the experimental observations without considering the
effect of the interactions of Bcd. In particular, taking into account
the distribution of Bcd-mRNA determined in~\cite{Little2011} we find
that the SDID model can account for the formation of the bulk part of
the Bcd gradient within the experimentally observed times. Although
the formation of the gradient is a nonlinear process that involves
several timescales, the analysis we present here shows that the {\it
  collective} effective coefficient, $D_{coll}$, gives a correct
estimate of the order of magnitude of the time it takes for $[Bcd]$ to
converge to its corresponding stationary distribution.

An important aspect of the \textit{bcd} morphogen system is the
precise response of one of its main target genes, \textit{hunchback}
(\textit{hb}). As well as \textit{bcd}, \textit{hb} is a maternal
effect gene. In the early embryo, the \textit{hb} mRNA is supplemented
with zygotically transcribed mRNA which production is regulated by
Bcd. The distribution of the resulting protein, Hb, presents very
sharp borders along the AP axis, as an ``on/off'' pattern. This
indicates a high sensitivity of the \textit{hb} mRNA to the
concentration of Bcd. In~\cite{Strhulet1989,Nusslein1989} 7 Bcd
binding sites in the \textit{hb} promoter were identified. Thus, the
idea of a cooperative Bcd binding was proposed to explain the sharp
borders of [Hb]. This implies that the concentrations of Hb and Bcd
are related by: 
\begin{equation}
\frac{[Hb]}{[Hb]_{max}}=\frac{[Bcd]^n}{[Bcd]^n+[Bcd]_{1/2}^n},
\label{eq:Hb_hill}
\end{equation}
where the Hill coefficient, $n$, accounts for the cooperativity and
$[Bcd]_{1/2}$ is the Bcd concentration at which [Hb] reaches half of
its maximun value. Using embryos immunostained for DNA, Bcd and Hb,
scatter plots of $[Hb]$ {\it vs.} $[Bcd]$ were obtained
in~\cite{Gregor:Cell2007b}.  From these plots a Hill coefficient,
$n=5$, was estimated. In spite of the relatively large value of $n$,
the scatter plots also showed a remarkable degree of precision between
the distributions of Hb and Bcd ($\sim 10\%$ near the point of
half-maximal activation)~\cite{Gregor:Cell2007b} in spite of the
fluctuations that are intrisic to the transcription
process~\cite{Little2013}.  The control experiments
of~\cite{Gregor:Cell2007a} showed that the fluorescence intensity
obtained in antibody stainings was linearly related to the one
collected from Bcd-GFP. The latter allows the observation of the
spatio-temporal dynamics of $[Bcd]$ {\it in vivo}. It is important to
remark that in experiments that use Bcd-GFP, it is not possible to
distinguish if Bcd is free or bound: the fluorescence distribution
profile corresponds to the total $[Bcd]$. This means that the actual
dynamics of the free Bcd is hidden in the observations. The length and
time scales of free Bcd could differ, in principle, from that of the
Bcd-GFP observed experimentally. In fact, in the present paper we show
that this is the case using the SDID model with realistic parameter
values. Assuming that the length-scale determined from the
observations of Bcd-GFP corresponds to that of free Bcd can lead to
incorrect estimations of the Bcd diffusion coefficient and/or
degradation rate.  The difference in the length-scale of the free and
the total $[Bcd]$, on the other hand, can have implications for the
relationship between $[Bcd]$ and $[Hb]$ that is derived from the
fluorescence observations.  As we show in the present paper, depending
on what the binding sites of the SDID model represent, the Hill
coefficient that characterizes the cooperativity with which Bcd
promotes the production of Hb can be different from the one that is
directly derived from the scatter plot of the observed
fluorescence. This, in turn, calls for a revision on the conclusions
about the precision with which hb reads the $[Bcd]$ distribution.  \\

%\subsection{Precise response to the Bcd gradient}

\section{Methods}

\subsection{The SDID model}
\label{sub:sdid}

We consider a model in which Bcd is synthesized over a region of the
anterior pole embracing 20\% of the embryo, diffuses with free
coefficient, $D_f$, interacts with a single type of binding sites
according to~(\ref{eq:reaction_scheme}) and is degraded at a
constant rate. We consider a cylindrical domain in which all
concentrations only vary along the axial coordinate, $z$, ($0\le z\le
L=500\mu m$). The assumptions on the Bcd synthesis are based on the
observations of~\cite{Little2011} according to which $\sim 90\%$ of
the mRNA of Bcd is located in a region that starts at the anterior
pole and extends up to $20\%$ of the length of the embryo. Although
the mRNA concentration in this anterior region is not perfectly
uniform, we simplify its description and assume that $Bcd$ is
synthesized with a rate:
\begin{equation}
\theta(z) = \left\{ 
\begin{array}{l l}
\theta_o &  \quad {\rm if} z\leq 0.2L\\
0 & \quad{\rm if} z>0.2L
\end{array} 
\right. ,
\end{equation}
where $\theta_o$ is constant. We assume that the binding sites are
uniformly distributed over the whole embryo and diffuse with
coefficient, $D_S\ll D_f$. We suppose that the binding sites belong to
molecules that are much more massive than Bcd so that their mobility
remains unaltered when they bind Bcd. In certain instances we also
assume that they are immobile.  We consider different alternative
versions of the model that differ in the manner in which we treat Bcd
degradation or in whether we include the dynamics of Bcd-GFP
maturation or not.  Here we present the equations of the version that
we call of ``partial degradation'' because we assume that Bcd is
degraded only in its free form. This version does not include the process of GFP maturation either.
The equations of the other versions of the SDID model
are described in the Appendix.  The spatio-temporal dynamics of the
concentrations if Bcd is degraded only when free and the process of
Bcd-GFP maturation is not included is given by:
\begin{equation}
\eqalign{
\frac{\partial{[Bcd]}}{\partial t} = & D_{f}\nabla^2 [Bcd] -k_{on}[Bcd]([S_T] - [Bcd_b]) +  k_{off}[Bcd_b] \\
&-\tilde{\alpha}[Bcd] + \theta(z),\\ 
\frac{\partial{[Bcd_b]}}{\partial t}  = & D_{S}\nabla^2 [Bcd_b] +k_{on}[Bcd]([S_T] - [Bcd_b]) - k_{off}[Bcd_b] ,
}
\label{eq:rd_eq}
\end{equation}
where $[Bcd_T] = [Bcd] + [Bcd_b]$ and $[S_T] = [S]+[Bcd_b]$ are the
total Bcd and binding site concentrations, respectively and
$\tilde{\alpha}$ is the degradation rate. To simplify the notation we
refer to the concentration of free Bcd as $[Bcd]$. It should not be
confused with Bcd, which generically refers to the protein. In this
model we assume that all Bcd, free or bound to sites, is fluorescent.

%\begin{eqnarray}
%\dot{[Bcd]} &=& D_{f}\nabla^2 [Bcd] -k_{on}[Bcd][S] + k_{off}[Bcd_b] -\alpha[Bcd] + \theta(z), \\
%\dot{[S]} & = & D_{S}\nabla^2 [S] -k_{on}[Bcd][S] + k_{off}[Bcd_b],\\
%\dot{[Bcd_b]} &= &D_{S}\nabla^2 [Bcd_b] + k_{on}[Bcd][S] - k_{off}[Bcd_b].
%\end{eqnarray}

\subsection{Analytical estimations: the SDID model under the fast reactions approximation} 

The {SDID} model in all of its versions is a 
reaction-diffusion system. The analysis of this kind of systems may
be complicated. Hereby, in order to interpret some results and choose
\textit{a priori} some parameters we introduce an approximation where
the Bcd transport is described in terms of an effective diffusion
coefficient.  If we consider that reactions take place on a much
faster time scale than diffusion a fast reaction~\cite{strier2000} or
\textit{fast buffering} approximation~\cite{Wagner:Biophysj1994} can
be used. Under this condition the systems given by
(\ref{eq:rd_eq}) (or (\ref{eq:rd_eq_deg_total})) can be
described as:
\begin{eqnarray}
%\eqalign{
\frac{\partial{[Bcd]}}{\partial t} &=& D_{coll}\nabla^2 [Bcd] -\beta D_S\vert\nabla[Bcd]\vert^2 -\hat{\alpha}[Bcd]  + \hat{\theta}(z,t),\label{eq:Bcd_fast_buffers_a}\\
 {[Bcd_b]} &=& \frac{[Bcd][S_T]}{[Bcd]+K_D},
%}
\label{eq:Bcd_fast_buffers_b}
\end{eqnarray}
where $D_{coll}$ is the effective (collective) diffusion coefficient
defined in~(\ref{eq:Dcoll}), $\hat{\theta}=\frac{\theta}{1 +
  \frac{[S]^2}{K_d [S_T]}}$, $\beta= \frac{2[S]^3}{K_d^2 [S_T]^2 (1 +
  \frac{[S]^2}{K_d [S_T]})}$ and $\hat{\alpha}=\frac{\tilde{\alpha}}{1
  + \frac{[S]^2}{K_d [S_T]}},$ for the partial degradation
model~(\ref{eq:rd_eq}), and
$\hat{\alpha}=\frac{\overline{\alpha}\left(1+\frac{[S]}{K_D}\right)}{1
  + \frac{[S]^2}{K_d [S_T]}}$ for the total degradation
model~(\ref{eq:rd_eq_deg_total}). Given that $D_S \ll D_f$, for
simplicity we will consider $D_S= 0$. In such a case the term $\propto
\vert \nabla [Bcd] \vert^2$ in~(\ref{eq:Bcd_fast_buffers_a}) can be
neglected and the first of the equations reduces to:
\begin{equation}
\frac{\partial{[Bcd]}}{\partial t} = D_{coll}\nabla^2 [Bcd] -\hat{\alpha}[Bcd] + \hat{\theta}(z).
\label{eq:Bcd_fast_buffers_Ds_0}
\end{equation}
Although this looks like a linear diffusion equation it is not, since
$D_{coll}$, $\hat{\alpha}$ and $\hat{\theta}$ depend on $[Bcd]$, {\it
  i.e.}, they are position and time dependent. The steady-state
solution of~(\ref{eq:Bcd_fast_buffers_Ds_0})  coincides with that of
(\ref{eq:rd_eq}) for $D_S=0$.

\subsection{Numerical simulations}

We solve the system of equations~(\ref{eq:rd_eq}),
(\ref{eq:rd_eq_deg_total}) and (\ref{eq:rd_eq_tagged_untagged}) using
the Douglas-Ratchford ADI method~\cite{Douglas-Rachford_ADI}. The
integration domain is a cylinder of length $L=500 \mu m$ with no flux
boundary conditions at both ends. We list in table~\ref{table_1} the
parameter values that we use. For the concentrations, dissociation
constant and free diffusion coefficients we use the estimates
presented in~\cite{Sigaut2014} which were derived from an analysis of
the experiments of~\cite{AbuArish:BJ2010}. We show in
section~\ref{sec:parameters} how the rest of the parameters were
chosen.

\subsection{Choice of parameter values}
We here describe how we choose the parameter values in the case of the
SDID model with partial degradation. To have a good starting point, we
first look at the stationary solution of (\ref{eq:rd_eq}) with
the goal of comparing semi-quantitatively the observed fluorescence
profile with the (stationary) distribution of the total Bcd
concentration, {\it i.e.}, of $[Bcd_T]=[Bcd_b]+[Bcd]$. To this end we
set $D_S=0$, a reasonable approximation given that $D_S\ll D_f$.
Assuming that the continuous production of Bcd eventually saturates
the binding sites inside the region where Bcd is synthesized ({\it
  i.e.}, for $z\le 0.2L$) we estimate that there is a subregion, $z\le
z_u\le 0.2L$, where $[Bcd]$ and $[Bcd_T]$ are approximately uniform
with $Bcd$ and $S$ in equilibrium between themselves:
\begin{eqnarray}
 {[Bcd_b]} &=& [Bcd][S_T]/([Bcd]+K_d),\label{eq:eq_Bcdb} \\
 {[S]} &=& K_d[S_T]/([Bcd]+K_d) , \label{eq:eq_S}
\end{eqnarray}
 and where there is a balance between the rate of Bcd production and
 degradation. Thus, we expect the stationary solution to satisfy:
\begin{eqnarray}
 {[Bcd]} &\approx & \frac{\theta}{\tilde{\alpha}}, \\
 {[Bcd_b]} &\approx & \frac{\theta}{\tilde{\alpha}}\frac{[S_T]}{\theta/\tilde{\alpha}+K_d} , \label{eq:max}
\end{eqnarray}
close to the anterior pole.  The observed Bcd concentration has been
estimated as $\sim 140nM$ at the anterior pole~\cite{AbuArish:BJ2010}.
Therefore, we choose $\theta/\tilde{\alpha}$, so that
$[Bcd_T]=[Bcd_b]+[Bcd]\approx 140 nm$ at $z=0$.  Outside the region of
Bcd synthesis ($0.2L \le z\le L$), the stationary solution also
satisfies the equilibrium condition
(\ref{eq:eq_Bcdb})--(\ref{eq:eq_S}).  The stationary distribution
of free Bcd in this region then satisfies
\begin{equation}
0=  D_{f}\nabla^2 [Bcd]_s  -\tilde{\alpha}[Bcd]_s . \label{eq:free_eq}
\end{equation}
Thus, it is given by~(\ref{eq:SDD}) with $z_o$ equal to:
\begin{equation}
z_{o_f}\equiv \sqrt{D_f/\tilde{\alpha}}. \label{eq:zof}
\end{equation}
$[Bcd_T]=[Bcd_b]+[Bcd]$ decays with a different lengthscale. Namely,
defining $z_{o_T} \equiv[Bcd_T]/\vert\nabla[Bcd_T]\vert$ (for $z>
0.2L$) and using (\ref{eq:SDD}) and (\ref{eq:eq_Bcdb}) we obtain
\begin{eqnarray}
\frac{z_{o_f}}{z_{o_T}} = \frac{D_{sm}}{D_{coll}},
\label{eq:z_o/z_oT}
\end{eqnarray}
where $D_{sm}$ and $D_{coll}$ are given by~(\ref{eq:Dsm})--(\ref{eq:Dcoll}) with $D_S=0$. Clearly, (\ref{eq:z_o/z_oT}) does not prescribe a single lengthscale,
$z_{o_T}$, since $D_{sm}$ and $D_{coll}$ depend on $[Bcd]$ which is
not uniform for $z> 0.2L$. However, using some ``typical''
concentration values along the gradient we find a first estimate of
$\tilde{\alpha}$. In particular, considering that $z_{oT}$  corresponds to the observed
characteristic lengthscale of the fluorescence gradient, $\ell_o\approx
(100-150)\mu m$, that $D_f \sim 20 \mu m^2/s$ and that $D_{sm}/D_{coll}\approx 0.1$ in
the region where FCS experiments are performed~\cite{Sigaut2014},
(\ref{eq:zof})--(\ref{eq:z_o/z_oT}) yield $\tilde{\alpha}\sim
(0.1-0.2)s^{-1}$.  Then, through the constraint that the total concentration
observed at the anterior pole imposes on $\theta/\tilde{\alpha}$
(\ref{eq:max}), we derive $\theta$. Based on these \textit{a
  priori} estimates we then explore the parameter space and choose
final values that are able to reproduce semi-quantitatively the
experimental observations.  In particular, we found that
$\tilde{\alpha}=0.05 s^{-1}$ allowed to reproduce most properties of
the observed gradient. Using this value of $\tilde{\alpha}$,
(\ref{eq:eq_S}) and $[Bcd_T](z=0)=140 nM$, we estimated
$\theta_0=0.5 nM s^{-1}$.

\begin{table}[tph]
\centering
%\resizebox{\columnwidth}{!}{%
%\resizebox{\linewidth}{!}{%
\begin{tabular}{c | c}
\hline
\hline
$D_S$ & $0.095 \mu m^2s^{-1}$\\
$D_f$ & $19 \mu m^2s^{-1}$\\
$k_{off}$ & $0.1 s^{-1}$ \\
$K_D$ & $0.192 nM$ \\
$[S_T]$ & $130 nM$\\
%$\gamma$ & $??$ \\ 
%$\alpha$ & $5.90 s^{-1}$\\
%$\theta_o$ & $68.29 \mu M s^{-1}$ \\
$L$ & $500 \mu m$\\
\hline
\hline
\end{tabular}
\caption{Common simulation parameters to all  SDID model versions. The rates
of Bcd degradation and synthesis and of GFP maturation used in the different
versions are given in the text.}
\label{table_1}
\end{table}

\section{Results and discussion}
%. 

\subsection{Reproduction of the gradient.} \label{sec:parameters}

Here we show the results obtained through numerical simulations of the
SDID model in its different versions and describe how the solutions
depend on the various parameters.  In the simulations we use the
values, $[S_T]$, $D_f$, $D_S$, $K_d$, and $[Bcd_T]$ (the latter, at
the anterior pole) derived in~\cite{Sigaut2010}. We use $L=500\mu
m$~\cite{Gregor:Cell2007a} and the rate, $\gamma$, at which Bcd-GFP
matures and becomes fluorescent estimated
in~\cite{Drocco2011,Liu2013}. The other parameters were chosen so that
the solutions reproduced semi-quantitatively the experimental
observations of~\cite{Gregor:Cell2007a,Little2011} as explained in
Methods and the Appendix.

\begin{figure}%[H]
   \begin{center}
      \includegraphics*[width=4.5in]{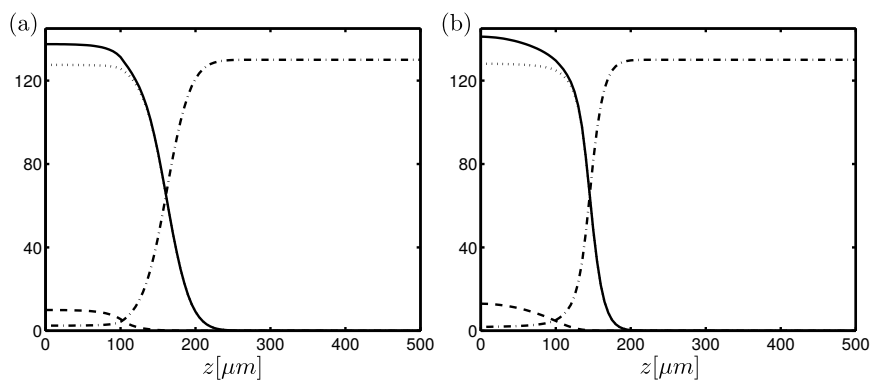}
\caption{Numerically simulated concentration distributions at $t=100$
  min and along the AP axis of $Bcd_T$ (solid curve), $Bcd$ (dashed
  curve), $Bcd_b$ (dotted curve) and $S$ (dashed-dotted curve). All
  concentrations are in $nM$. (a) Results of the model with partial
  degradation, $\tilde{\alpha}=0.05s^{-1}$ and $\theta_0=0.5 nM
  s^{-1}$. (b) Results of the model with total degradation,
  $\overline{\alpha}=0.0005s^{-1}$ and $\overline{\theta}_0=0.1 nM
  s^{-1}$.}
      \label{fig:figure1}
   \end{center}
\end{figure}

We show in figure~\ref{fig:figure1} the concentration of the different
species along the AP axis obtained at time, $t=100$ min, using the
partial~(a) and total degradation models~(b) with the parameters of
table~\ref{table_1} and the others as described in the caption.  For
both models we see that the Bcd$_T$ profile is consistent with the
experimental observations. In particular, its concentration decays to
$\sim 50 \%$ of its maximum value at $z=150 mu m$. We also observe
that most Bcd molecules are bound at the anterior pole (as estimated
in~\cite{Sigaut2014}). As we move away from the anterior pole,
$[Bcd_T]$ begins to decrease and the number of free binding sites
increase. Since the mobility of the bound molecules is slower than
that of the free ones, the gradient of $[Bcd_b]$, and hence of
$[Bcd_T]$, is more extended than that of $[Bcd]$.  Comparing
figures~\ref{fig:figure1}~(a) and (b) we observe that the Bcd$_T$
distribution is slightly different due to the difference in how the
Bcd degradation is treated. In particular, near the anterior pole the
slope of $[Bcd_T]$ is more pronounced in the total degradation
case. This agrees with a small difference in the lengthscale of the
gradient obtained with each model. In the case of the partial
degradation model, $[Bcd_T]$ decays to $10\%$ of its maximum at
$z\sim 195\mu m$, while in the case of the total degradation model the
same percent is achieved at $z\sim 165 \mu m$. 
figure~\ref{fig:figure1} also shows that in neither model the free or
the total Bcd concentrations follow the mRNA distribution given by
$\theta (z)$. This implies that, for our model,  the Bcd concentration distribution is
not a passive reflection of the mRNA that produces it, as assumed
in~\cite{Spirov2009}.

Comparing the values of $\tilde{\alpha}$ and $\overline{\alpha}$ used
in figure~\ref{fig:figure1} with previously reported ones, $0.0003
s^{-1} - 0.0015 s^{-1}$~\cite{Grimm2010,Gregor:Cell2007a,Drocco2011},
we observe that the one used in the total degradation model is within
this previous estimated range, while the one used in the partial
degradation model is larger. This discrepancy is reasonable if we take
into account that degradation rates were estimated from fluorescence
records that did not distinguish between bound and free Bcd. Namely, if
only free Bcd is degraded with rate $\tilde{\alpha}$, ({\it i.e.}, $\partial [Bcd]/\partial t \sim -\alpha [Bcd]$), the rate at which
$[Bcd_T]$ decreases due to degradation is given by $\tilde{\alpha}[Bcd]/[Bcd]_T$ ({\it i.e.}, $\partial [Bcd_T]/\partial t \sim -\alpha [Bcd]/[Bcd]_T\,  [Bcd]_T$). Degradation rates determined from fluorescence images
would then correspond to $\tilde{\alpha}[Bcd]/[Bcd]_T$.
For the parameter values of
figure~\ref{fig:figure1}(a), this fraction it is $[Bcd]/[Bcd]_T\approx\sim 0.08$, which allows
to explain in part the discrepancy between the degradation rate used,
$\tilde{\alpha}=0.05s^{-1}$, and the values reported previously in the
literature~\cite{Grimm2010,Gregor:Cell2007a,Drocco2011}.

Regarding the rate of protein synthesis, $\theta$, the number of mRNA
molecules inside the embryo during n.c. 10-13 was estimated as $\sim
10^5$ molecules~\cite{Little2011}. Considering that mRNA molecules are
mainly synthesized over a region that occupies $\sim 20$\% of the
total length of the embyro, that each molecule can synthesize one
protein per second~\cite{Bionumbers}, that the region where nuclei are
located and Bcd diffuses is the $20 \mu m$-wide outermost ``slice'' 
of the embryo and approximating the latter by an ellipsoid of radii $\sim
250\mu m$, $75 \mu m$ and $75 \mu m$, the $ 10^5$ mRNA molecules
imply  a $\sim
0.3 nM/s$ Bcd synthesis rate. This value is similar to the ones used
in the SDID model with partial or total degradation (see figure~\ref{fig:figure1}).

\begin{figure}%[H]
   \begin{center}
      \includegraphics*[width=4.5in]{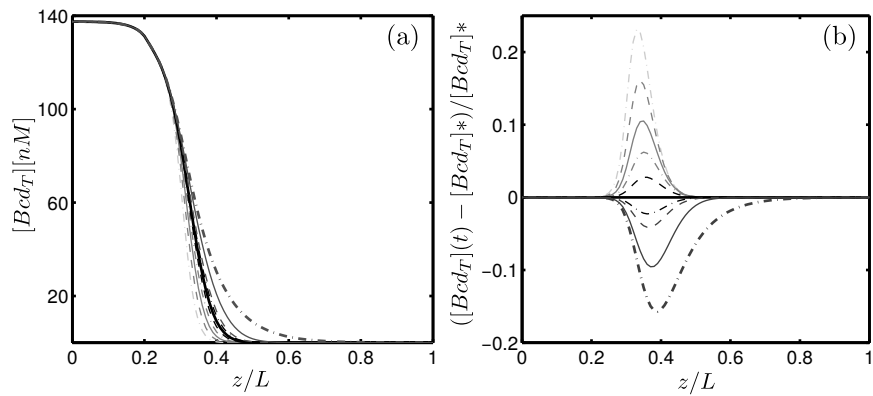}
\caption{(a) Distribution of total Bcd along the AP axis obtained at, increasing from left to right, 
  $t=50$, 60, 70,
    80, 90, 100, 110, 120, 167, 833 min, using the model with partial degradation and the parameters of figure~\ref{fig:figure1}~(a). The solid curve at the center corresponds to $t=100$ min, the time at which the gradient is supposedly already
  established. (b) Differences between $[Bcd_T]$ at different times
  and $[Bcd_T]*\equiv [Bcd_T](t=100min)$. For $t\in (70, 167)$ min,
  $[Bcd_T]$ remains almost constant along the AP axis, except for a
  small region where the differences are less than 10\%.}
      \label{fig:figure2}
   \end{center}
\end{figure}

So far we have discussed the spatial properties of the gradient. The
time it takes for the gradient to be formed is another important
aspect that has been debated at large, mainly because the diffusion
coefficient estimated in~\cite{Gregor:Cell2007a} was too small to
account for the gradient formation within the experimentally observed
times.  Changing the reaction rate, $k_{off}$, while keeping $K_D$
constant it is possible to modify the timescale over which the
reactions take place.  We thus explored the predictions of the SDID
model using the parameters of figure~\ref{fig:figure1} varying $k_{off}$
over the range $(10^{-5}-10^3)s^{-1}$. We found that the results were
mostly insensitive to variations in $k_{off}$ (data not shown). We
observed that the concentration distributions only changed for the
most extreme (unrealistic) values of $k_{off}$. We thus chose
$k_{off}=0.1s^{-10}$, which we think is reasonable for the type of
interactions that Bcd may experience. We show in
figure~\ref{fig:figure2}(a) the distribution of the total Bcd
concentration at various times obtained using the partial degradation
SDID model. There we can observe how $[Bcd]_T$ converges to its
asymptotic value with increasing time. In particular, the time it
takes for it to be within $5\%$ of the stationary solution is more
than $t=800$ min ($\approx 13$ h). This time is much larger than the
100 min observed in experiments~\cite{Gregor:Cell2007a}. However, for
$t\in (70,167)$min the distribution does not change significantly, as
illustrated in figure~\ref{fig:figure2}(b). During this time interval
$[Bcd_T]$ differs by less than $10\%$ with respect to the
concentration at $t=100$ min and the largest differences are
restricted to a very small spatial region. Moreover, the difference
with respect to the stationary distribution for $t\ge 70$ min is never
larger than $\sim 15\%$ regardless of position.  $15\%$ differences
are in the border of experimental detectability (particularly, far
away from the anterior pole).  Our results then suggest, in accordance
with the work of \textit{Bergmann et al}~\cite{Bergmann2007}, that the
steady-state is not reached in less than $100$ min but that yet the
gradient may seem stationary. In the case of the SDID model with total
degradation we observe a similar evolution of $[Bcd_T]$. For $t\in
(80,120)$ min, $[Bcd_T]$ varies by less than $20\%$. Although the
rates of production and degradation for the partial and total
degradation models are different, the spatio-temporal dynamics of the
free and total Bcd concentrations are similar. For this reason, from
now on we will present results corresponding to the partial
degradation model only.

\begin{figure}[ht]
   \begin{center}
      \includegraphics*[width=4.5in]{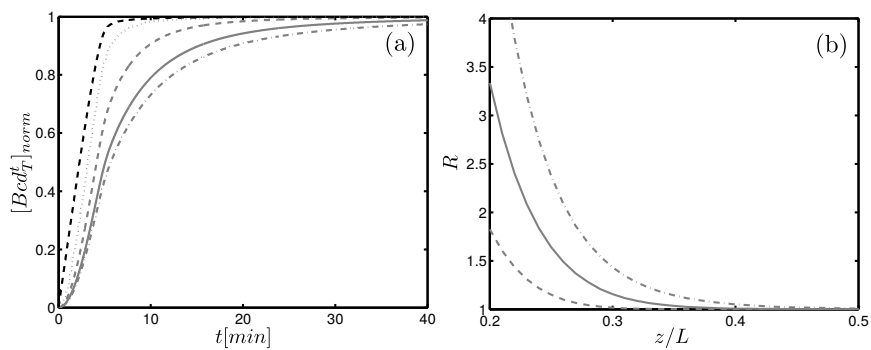}
\caption{(a) Time dependence of the total fluorescent Bcd
  concentration at the position where the source ends ($z=20\mu m$)
  obtained through simulations of
  (\ref{eq:rd_eq_tagged_untagged}) for different rates of GFP
  maturation (gray dotted line: $\gamma=0.01 s^{-1}$, gray dashed
  line: $\gamma = 0.003 s^{-1}$, gray solid line: $\gamma = 0.001
  s^{-1}$ and gray dashed-dotted line: $\gamma = 0.0005 s^{-1}$). The
  total Bcd concentration obtained without including the process of
  GFP maturation is shown with a black dashed curve. All parameters
  are the same as in figure~\ref{fig:figure1}(a) and the concentrations
  displayed are normalized by their corresponding asymptotic
  values. (b) Ratio of fluorescent to total Bcd, $R$, along the
  anterior-posterior axis at time, $t=100 min$ for three of the
  maturation rates probed in (a). Symbols are the same as in (a)}
\label{fig:figure3} \end{center}
\end{figure}

There is still one property of our simulations that is incompatible
with the observations: the maximum concentration ($[Bcd_T](z=0)$) is
reached almost instantaneously. This does not agree with the results
of~\cite{Little2011} where it is observed that the maximum is reached
$80$ minutes after fertilization. This discrepancy might be due to the
finite time it takes for GFP to mature and become fluorescent. To
evaluate the effect of maturation we performed numerical simulations
of the system given by (\ref{eq:rd_eq_tagged_untagged}). We show
in figure~\ref{fig:figure3}(a) the total Bcd concentration normalized by
its asymptotic value at $z=100 \mu m$ as a function of time for
different maturation rates, $\gamma$. As a reference we also show the
corresponding curve for the SDID model with no maturation (black
dashed curve). The parameters used in the simulations are the same as
in figure~\ref{fig:figure1}(a). As expected, the convergence to the
asymptotic value takes longer for the model that incorporates
maturation and becomes slower as $\gamma$ decreases. For $\gamma =
0.01 s^{-1}$ we observe little differences between the models with and
without maturation. At $t=10$ min the concentration reaches $98.1\%$
of the asymptotic value in the model with maturation and $99.5\%$ in
the model without maturation. In the case with $\gamma = 0.0005
s^{-1}$ the concentration reached at $t=10$ min is $73\%$ of the
asymptotic value. Hence, for low values of $\gamma$ a significant
delay in the convergence is observed in the region close or at the
source. The delays obtained, however, never exceeded $\sim 50\%$ of the time
elapsed at very early times and
this gap decreased rapidly as time went by.  In regions far from the source,
the fraction of mature to total Bcd-GFP molecules is larger because it
takes longer for the molecules to reach those regions and in that time
they mature and become fluorescent. These results are similar
regardless of whether we consider the SDID model with partial or total
Bcd degradation.  This disparity in the delay to reach steady-state
depending on the position along the AP axis affects the fluorescence
spatial distribution implying a change in the relationship between the
lengthscale of the observed gradient and the parameters of the
model~\cite{Liu2013}.

To analyze the effect of maturation on the lengthscale we show in
figure~\ref{fig:figure3}(b) the ratio between total Bcd over fluorescent
Bcd, $R(z)\equiv [Bcd_T]/([Bcd^t]+[Bcd_b^t])$. $R$ depends on the
relation between $\tilde{\alpha}$ and $\gamma$. In
figure~\ref{fig:figure3}(b) we show the value, $R$, as a function of $z$
for different values of $\gamma$ ($[0.003-0.001-0.0005]
s^{-1}$). For all cases $R$ is larger near the source and decreases to
$\sim 1$ as the posterior pole is reached. In~\cite{Liu2013} it was
determined that $R \sim 3$ close to the source. Of all the considered
values of $\gamma$ the one that gives the most similar result to this
observation is $\gamma = 0.001 s^{-1}$. This value of $\gamma$ is
perfect agreement with the rate of degradation of Bcd reported
in~\cite{Grimm2010,Gregor:Cell2007a,Drocco2011}. The fact that
$R$ is a decreasing function of the distance to the source also changes
the lengthscale of the gradient with respect to the case in which 
GFP maturation is not considered. In particular, if we compare the
distribution, $[Bcd_T]$, for the same parameters as in figure~\ref{fig:figure3}
with and without including the process of GFP maturation the characteristic
lengthscales differ by a factor of 3.

\subsection{Interpretation of the experimental observations with a model that includes reactions.}

The results of the simulations presented so far show that it is
possible to reproduce the spatio-temporal characteristics of the
Bicoid gradient semi-quantitatively using the SDID model with
``reasonable'' biophysical parameter values. We now discuss how the
experimentally observed properties are related to the parameters of
the model.  More specifically, we are interested in determining the
relationship between these parameters and the length and timescales of
the Bcd gradient and how these relationships change depending on
whether the reactions with binding sites are included in the model or
not. Thus, we are after a re-interpretation of the observations within
the framework of a model that includes reactions. Such a model is
nonlinear and the concentrations are not characterized by a single
spatial or temporal scale. This becomes evident in the fast reaction
approximation, (\ref{eq:Bcd_fast_buffers_Ds_0}), where the transport
rate is determined by an effective diffusion coefficient which depends
on the concentration, and hence, does not have a single value along
the embryo or over time. The presence of ``many'' coefficients or
``multiple'' scales is in agreement with the work of~\cite{Little2011}
in which, in order to reproduce the experimental observations, a model
with diffusion coefficients that changed in time {\it ad hoc} was
introduced. This reinforces our idea that it is the effective
diffusion coefficients, which naturally arise within the context of
the SDID model, that determine the characteristic scales of the
problem.  Here we seek to relate the effective parameters of the model
with the observed spatial scale and the convergence time of the
gradient in the simplest possible way.  To this end we work with the
SDID model with partial degradation and $D_S=0$. The difference with respect to
the SDID model with total degradation is mainly a matter of parameter
values.  The differences with respect to the model that includes the
delay in GFP maturation is discussed later.

\subsubsection{Lengthscale: free Bcd gradient {\it vs} total Bcd gradient}

\begin{figure}[h]
   \begin{center}
      \includegraphics*[width=4.5in]{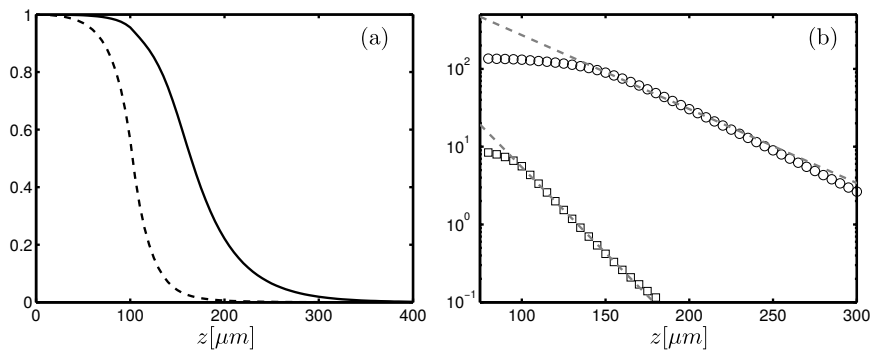}
\caption{(a) Normalized $[Bcd_T]$ (solid line) and $[Bcd]$ (dashed
  line) as functions of $z$ at time, $t=100 min$, obtained from
  numerical simulations of the SDID model with partial
  degradation and the same parameters as in figure~\ref{fig:figure1}~(a). (b) Similar to (a), with the vertical axis on a
  logarithmic scale ($[Bcd_T]$ with $\bigcirc$ and $[Bcd]$ with
  $\square$). Exponential fits to the distributions over the regions
  $z=(125-300)\mu m$ for $[Bcd]$ and $z=(150-300)\mu m$ for $[Bcd_T]$
  are superimposed. The characteristic lengthscales obtained with the
  fits are, respectively, $z_{of}\approx 19 \mu m$ and $z_{oT}\approx
  45 \mu m$.}
      \label{fig:figure4}
   \end{center}
\end{figure}

As discussed in the Introduction, the stationary solution of the SDD
model with a source at one end is given by
(\ref{eq:SDD})--(\ref{eq:long_car_sdd}). Within the framework of
this model $z_o$ corresponds to the characteristic lengthscale,
$\ell_o$, of the observed fluorescence distribution. As described in
the Methods Section, the stationary solution of the SDID model with
partial degradation satisfies
(\ref{eq:Bcd_fast_buffers_b})--(\ref{eq:Bcd_fast_buffers_Ds_0}).
Within the context of this model, the lengthscale of the free Bcd
concentration, $z_{o_f}$, is given by (\ref{eq:zof}) (and
(\ref{eq:zof_deg_tot}) in the case of the SDID model with total
degradation).  This lengthscale does not correspond to that of the
observed gradient because the fluorescence cannot distinguish between
free and bound Bcd. Therefore, the observed lengthscale, $\ell_o$,
should be related to that of the total Bcd concentration, $z_{o_T}$,
given by (\ref{eq:z_o/z_oT}).  $z_{o_T}$ and $z_{o_f}$ can be very
different between themselves. Moreover, $z_{o_T}$ changes with
position and time.
%%%%%%%%
%If the value $D \approx 1\mu^2 m/s$ estimated
%in~\cite{Gregor:Cell2007a,AbuArish:BJ2010} using FRAP is considered we
%obtained $\alpha\sim 0.0001s^{-1}$ while $\alpha\sim 0.002s^{-1}$ is
%obtained if it is considered the free diffusion coefficient $D\approx
%20\mu^2 m/s$ estimated in~\cite{Sigaut2014}. Both values are
%approximately within the range reported in the literature $0.0003
%s^{-1}- 0.0015
%s^{-1}$~\cite{Grimm2010,Gregor:Cell2007a,Drocco2011}. However, it is
%possible to work with the reduced equations in the approximation of
%fast
%reactions~(\ref{eq:Bcd_fast_buffers_b})--(\ref{eq:Bcd_fast_buffers_Ds_0})
%as was done for the estimation of parameters {\it a priori}. 
%%%%%%%%
We now discuss in what regions (\ref{eq:z_o/z_oT}) provides a good
estimate of the characteristic lengthscale of the total Bcd
distribution.  We show in figure~\ref{fig:figure4}(a) the normalized
free and total Bcd concentrations as functions of $z$ at $ t = 100 min
$. The free Bcd distribution decays by 50\% for $z \approx 100 \mu m$,
while the total Bcd concentration does it at $z \approx 160 \mu m$.
Although the total concentration does not decay exactly exponentially
with $z$ as in (\ref{eq:SDD}), it can be approximated by such an
expression over a certain range of $z$ values.  Free Bcd does follow
an exponential decay over the region where there is no source.  This
can be observed in figure~\ref{fig:figure4}(b) where we plot $[Bcd_T]$
and $[Bcd]$, normalized by their maximum values, as functions of
$z$. The exponential fits (linear on the logarithmic scale of the
figure) were done over the regions $z=(125-300)\mu m$ for $[Bcd]$ and
$z=(150-300)\mu m$ for $[Bcd_T]$ obtaining $z_{of}\approx 19 \mu m$
and $z_{oT} \approx 45\mu m$. These values can be compared with those
predicted by (\ref{eq:zof})--(\ref{eq:z_o/z_oT}). In the case
of free Bcd, the characteristic lengthscale given by
(\ref{eq:zof}) with the simulation parameters is $z_{of} = 19.5
\mu m $ which agrees with the fitted value.  In the case of $[Bcd_T]$
the comparison is more complicated because the lengthscale of
(\ref{eq:z_o/z_oT}) depends on $D_{coll}$ and $D_{sm}$ which vary
with time and space.  If we consider the values, $D_{coll}$ and
$D_{sm}$ at time $t=100$ min and over the region where the fitting
begins, $z=150\mu m$, we obtain $D_{coll}/D_{sm} \approx 2$. Inserting
those values in (\ref{eq:z_o/z_oT}) we obtain $z_{oT} \approx 40
\mu m $ which is very similar to the one estimated from the
fitting. If instead we consider the values at $z=180\mu m$, the ratio
of effective coefficients is $D_{coll}/D_{sm}\approx 1.2$, leading to
an estimate of $z_{oT} \approx 23 \mu m$ which only differs by a
factor of 2 with respect to the fitted value. Thus,
(\ref{eq:zof}) and (\ref{eq:z_o/z_oT}) provide good estimates of
the characteristic lengthscales of the free and total Bcd
concentrations if we use the values of the effective coefficients in
the region just contiguous to the source (where concentrations start
to decrease). We obtain similar results using
(\ref{eq:zof_deg_tot}) and (\ref{eq:z_o/z_oT_deg_tot}) within the
framework of the SDID model with total degradation.

\subsubsection{Timescale: effective {\it vs.} free diffusion coefficients}

The SDD model given by (\ref{eq:ec_sdd})--(\ref{eq:ec_sdd_bc}),
for long enough $L$, has solutions of the form~\cite{Bergmann2007}:
\begin{equation}
\eqalign{
{[Bcd]}(z,t) =& \frac{\zeta}{A\sqrt{\alpha D}}\left(\exp(-z/z_o)-\frac{\exp(-z/z_o)}{2} {\rm erfc}\left(\frac{2Dt/z_o-z}{\sqrt{4Dt}}\right) \right.\\
&\left. -\frac{\exp(z/z_o)}{2} {\rm erfc}\left(\frac{2Dt/z_o+z}{\sqrt{4Dt}}\right)\right),}
\label{eq:bergmann}
\end{equation}
where erfc  is the complementary error function, erfc($z)=1-\int_0^z\exp(-t^2)dt/\sqrt{\pi}$.
 This equation shows that the approach to the stationary solution (\ref{eq:SDD}) occurs as if there was a front that travels at speed: 
\begin{equation}
v\equiv 2\frac{D}{z_o}=2\sqrt{\alpha D} ,\label{eq:vel}
\end{equation}
that depends on the diffusion coefficient $D$ and the rate of degradation $\alpha$, and allows to define a convergence time at a distance $z$ from the source as:
\begin{equation}
t_{conv}(z)\equiv \frac{z}{\sqrt{\alpha D}} .\label{eq:t_conv}
\end{equation}
Thus, if $D$ is known {\it a priori} and $\alpha$ is chosen so that the theoretical
characteristic lengthscale of (\ref{eq:long_car_sdd}) corresponds to the observed fluorescence lengthscale, $\ell_o\sim 100\mu m$,  the convergence time can be rewritten as:
\begin{equation}
t_{conv}(z)\equiv \frac{z \ell_o}{D} .\label{eq:t_conv_derived}
\end{equation}
At $z=0.75L$, this time is too long ($\ge 10 hs$) if it is assumed
that $D\approx 1\mu^2 m/s$, the value estimated
in~\cite{Gregor:Cell2007a,AbuArish:BJ2010} using FRAP, and it is too
short ($\sim 0.5hs$) if the free diffusion coefficient of Bcd
estimated in~\cite{Sigaut2014}, $D\approx 20 \mu^2m/s$, is used
instead. The solution~(\ref{eq:bergmann}) also shows that the rate of
production, $\theta_o$, determines the maximum value of $[Bcd]$ but
is not involved in the convergence time.

In the case of the SDID model it is more difficult to define a
``propagation speed'' because in addition to the characteristic Bcd
degradation time there are other timescales related to the
reaction. In order to derive a propagation speed in such a case we
then work with the reduced equations in the fast reaction
approximation ((\ref{eq:Bcd_fast_buffers_Ds_0}) and (\ref{eq:Bcd_fast_buffers_b})). Given the formal
equivalence between (\ref{eq:Bcd_fast_buffers_Ds_0}) and
(\ref{eq:ec_sdd})-(\ref{eq:ec_sdd_bc}) we define the speed and the convergence time as in
(\ref{eq:vel}) and~(\ref{eq:t_conv}) but using $D_{coll}$ instead
of $D$ and $\hat{\alpha}$ instead of $\alpha$. We obtain:
\begin{equation} 
v = \sqrt{\tilde{\alpha} \frac{D_{coll}^2}{D_f}}
,\quad t_{conv}(z)= z\left(\frac{D_f}{\tilde{\alpha}
D_{coll}^2}\right)^{1/2},\label{eq:v_t_parcial}
\end{equation}
for the model with partial degradation and
\begin{equation}
\eqalign{
v = \sqrt{\overline{\alpha} \frac{D_{coll}^2}{D_{sm}}}, \quad t_{conv}(z)= z\left(\frac{D_{sm}}{\overline{\alpha} D_{coll}^2}\right)^{1/2},}
\label{eq:v_t_total}
\end{equation}
for the model with total degradation. As we did for the lengthscale,
we now analyze whether the solution obtained numerically for the model
with partial degradation moves with the speed described by
(\ref{eq:v_t_parcial}), if it is possible to define a single
characteristic value for the speed in the region immediately adjacent
to the source and, in that case, which effective diffusion
coefficients determine it. As in the case of the SDD model, if the
value, $\tilde{\alpha}$, is determined setting $z_{o_T}=\ell_o$ with
$\ell_o$ the observed fluorescence lengthscale and $z_{o_T}$ given by
(\ref{eq:z_o/z_oT})--(\ref{eq:zof}) with known values of $D_f$
qne $D_{sm}/D_{coll}$, the convergence time can be
rewritten as:
\begin{equation} 
t_{conv}=\frac{D_{sm} \ell_o z}{D_{coll}^2} . \label{eq:t_conv_gen}
\end{equation}
The same expression is obtained for the model with total degradation
using (\ref{eq:z_o/z_oT_deg_tot}) and (\ref{eq:zof_deg_tot}) and
setting $z'_{o_T}=\ell_o$. It then follows that if the degradation
rate, $\tilde{\alpha}$ or $\overline{\alpha}$, is derived from the
observed fluorescence lengthscale, the convergence time to the steady
state solution will be the same regardless of whether we use the model
with total or partial degradation. We now continue the analysis for
the model with partial degradation.

\begin{figure}[h]
   \begin{center}
      \includegraphics*[width=2.5in]{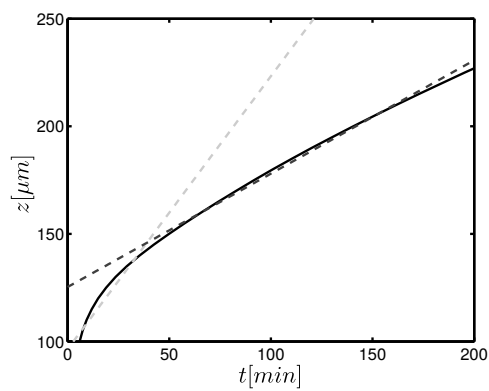}
\caption{Position along the AP axis and time, $(t,z)$, at which the
  free Bcd concentration reaches 50\% of its asymptotic value at the
  same $z$ in the case of the model with partial degradation (solid
  line) and the same parameters of figure~\ref{fig:figure1}~(a). To estimate the rate at which the gradient converges to its
  steady state two linear fittings were done in different regions
  along the AP axis (dashed lines).}
      \label{fig:figure5}
   \end{center}
\end{figure}

In order to represent the advancement of the Bcd front and
characterize its timescale we compute for each position, $z$, the
time, $t$, at which the free Bcd concentration, $[Bcd(z,t)]$, reaches
50\% of its asymptotic maximum value, $\max_t[Bcd(z,t)]$. We plot in
figure~\ref{fig:figure5} the position, $z$, $vs$ the time, $t$, just
defined. The slope of this curve corresponds to the propagation
speed. As expected in this case the front does not move with a
constant speed. It can be observed that the speed is smaller the
smaller $z$ is. As for the analysis of the lengthscale, here we focus
on a region where the speed is approximately constant. Based on the
results of figure~\ref{fig:figure5} we fit the front profile with a
linear function in two regions: $z\in [100,150]\mu m$ (light dashed
line) and $z\in[150,225]\mu m$ (dark dashed line). From the fits we
determine the speeds $v \approx 1.3 \mu m/min$ in the region closest
to the source and $v \approx 0.53 \mu m/min$ in the region further
away. This implies that at a distance of the source of the order of
the observed fluorescence lengthscale, $z\sim (100-140)\mu m$, the
convergence time is of the order of 77-100$min$, similar to the
characteristic time of the gradient formation obtained experimentally
($\sim 90\min$). We must point out that even if we here analyze the
convergence of the free Bcd concentration to its steady state
solution, the total Bcd (free and bound) reaches its asymptotic
distribution on a similar timescale.

We now analyze whether there is a simple expression that can be used
to estimate the speed and the convergence time. To this end we compare
the speed estimates of figure~\ref{fig:figure5} with those predicted by
(\ref{eq:v_t_parcial}). The latter gives $v(z=140\mu m, 50min)
\approx 1.5 \mu m/min$ and $v(z=180\mu m, 100min) \approx 0.4 \mu
m/min$. These estimates are similar to those derived with the
fitting. We then conclude that it is possible to relate the model
parameters with the timescale of the Bcd gradient formation in a
relatively simple way. (\ref{eq:v_t_parcial}) also highlights the
importance of distinguishing between the collective and single
molecule diffusion coefficients. If in (\ref{eq:v_t_parcial}) we
replace $D_{coll}$ by $D_{sm}$ the estimates of the front velocity
decrease by approximately one half implying that the timescale would be
twice the value derived before.

\subsubsection{The distinction between free and total Bcd and the role of Bcd as a transcription factor.}

As we have already mentioned, Bcd acts as a transcription factor for
the expression of hb. This process has been studied in detail both
experimentally and
theoretically~\cite{Gregor:Cell2007b,Tkacik26082008,Dubuis08102013}. In
particular, the observed (fluorescence) distributions coming from Bcd and Hb in fixed embryos
have been used to develop the theory. These distributions  were found to be related by
a non-linear function with Hill coefficient $n=5$, {\it i.e.},
consistent with a high degree of cooperativity of Bcd for the
transcription of hb~\cite{Gregor:Cell2007b}. These observations, however,
cannot distinguish between free and bound Bcd. Here we explore how these
observations should be re-interpreted when the interaction of Bcd with
binding sites is taken into account. To this end we consider two
possible scenarios. In one of them we assume that the sites, $S$, of
the scheme (\ref{eq:reaction_scheme}), correspond to specific sites
for the transcription of hb.  In the other situation we assume that
$S$ represents other binding sites either on DNA or in other species
({\it e.g.} RNA) and that the binding of Bcd to the promoters of the
hb transcription does not alter significantly the concentrations of
free or bound Bcd that enter the scheme
(\ref{eq:reaction_scheme}). This last scenario fits nicely within the
sliding and hopping model in which trasncription factors bind to
non-specific sites and then eventually find the specific sites for
transcription on the DNA molecule~\cite{vonHippel15011989}.  In
particular, the binding to DNA with different affinities, probably
associated to specific and non-specific sites, has recently been
quantified in mouse embryos~\cite{White201675}.  
 
\begin{figure}[h]
   \begin{center}
      \includegraphics*[width=4.25in]{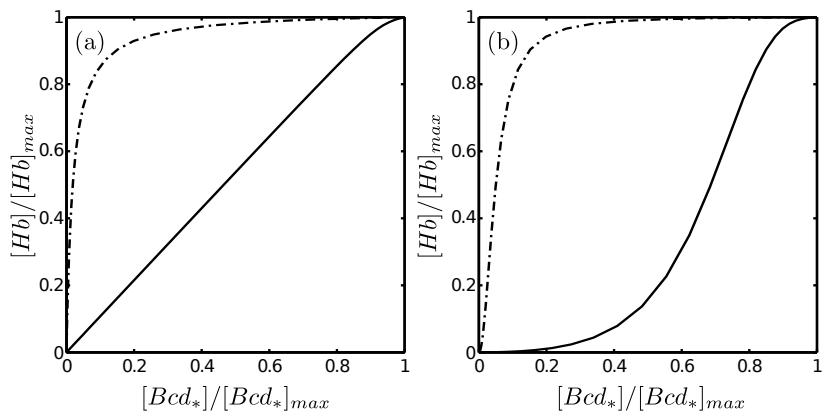}
\caption{Concentration of Hb, $[Hb]$, as a function of the normalized
  concentrations of free (solid line) and total (dashed line) Bcd that
  correspond to the steady state solution of the SDID model with
  partial degradation and the parameters of table~\ref{table_1}. (a)
  In this case $S$ corresponds to the binding sites that are specific
  for the transcription of hb so that $[Hb]\propto [Bcd_b]$. (b) In
  this case $S$ corresponds to non-specific binding sites that compete
  for Bcd with those that promote the transcription of hb. [Hb] is
  computed as in (\ref{eq:Hb_hill}) with $n=2$ and
  $[Bcd]_{1/2}=\overline K_D = 0.5\mu M$.}
      \label{fig:figure6}
   \end{center}
\end{figure}

We first analyze the scenario in which the sites that interact with
Bcd are the promoters for the synthesis of the mRNA involved in the
production of Hb. For simplicity, we here assume that the
concentration of Hb is proportional to the concentration of bound Bcd,
$[Bcd_b]$. This implies a simplification because the SDID model
considered here does not include a cooperative scheme for the
interaction of Bcd and the binding sites, S (see
(\ref{eq:reaction_scheme})). In any case, the purpose of this
analysis is to estimate by how much the Hill coefficient that can be
inferred from the observations could vary if it is assumed that the
observed fluorescence is proportional to the free or the total Bcd
concentrations. In this first scenario the concentrations are related
by:
\begin{equation}
\frac{[Hb]}{[Hb]_{max}}=\frac{[Bcd_b]}{[Bcd_b]_{max}} = \frac{[Bcd]}{[Bcd] + K_D} ,
\label{eq:Hb_es_Bcdb}
\end{equation}
with $K_D$ the dissociation constant of the reaction scheme
(\ref{eq:reaction_scheme}). We show in figure~\ref{fig:figure6}(a)
$[Hb]/[Hb]_{max}$, computed as prescribed by
(\ref{eq:Hb_es_Bcdb}), as a function of the free ($[Bcd_*]=[Bcd]$)
and total ($[Bcd_*]=[Bcd_T]$) Bcd concentrations normalized by their
maximum values. The figure reflects the fact that most Bcd is bound
so that $[Bcd_b]\approx [Bcd_T]$.  A similar plot would have been
obtained within this scenario if most Bcd were bound regardless of the
cooperativity of the interaction. The relationship between
$[Hb]/[Hb]_{max}$ and $[Bcd_T]/[Bcd_T]_{max}$ is very different from
the observations of~\cite{Gregor:Cell2007b} that show a nonlinear
relationship between the fluorescence coming from Hb and Bcd. As may
be observed in the figure, the relationship between $[Hb]/[Hb]_{max}$
and $[Bcd]/[Bcd]_{max}$ is very nonlinear. However, the range of
$[Bcd]$ values for which [Hb] is most sensitive to changes in $[Bcd]$
does not agree with the observations. These results suggest that Bcd
not only interacts with the specific DNA sites that promote the
production of Hb.

\begin{figure}[h]
   \begin{center}
      \includegraphics*[width=4.25in]{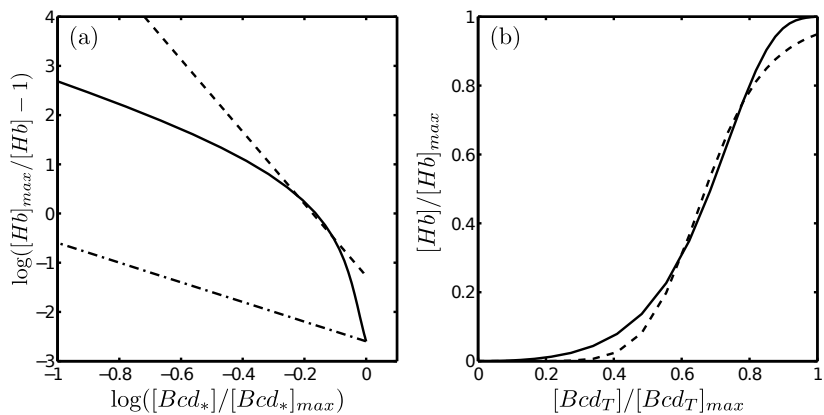}
\caption{(a) $\log({[Hb]_{max}}/{[Hb]}-1)$ $vs$
  $\log({[Bcd_*]}{[Bcd_*]_{max}})$ for $Bcd_* = Bcd$ (dashed-dotted line) and
  $Bcd_*=Bcd_T$ (solid line), with the various concentrations
  computed as in figure~\ref{fig:figure6}. The dashed curve corresponds
  to a linear fit for the case with $Bcd_*=Bcd_T$ in the region of
  greatest $[Hb]$ sensitivity to $Bcd_T$ variations. The Hill
  coefficient estimated from this fitting is $n_{f} \approx 7$. (b)
  Normalized $[Hb]$ as a function of $[Bcd_T]$ (solid curve) and the relationship prescribed by (\ref{eq:Hb_hill_nf}) with $Bcd_*=Bcd_T$ and $n_f=7$ (dashed curve).}
      \label{fig:figure7}
   \end{center}
\end{figure}

We now analyze the second scenario in which
(\ref{eq:reaction_scheme}) represents the interaction with
non-specific binding sites. For the results obtained in the previous
Sections to be applicable within this scenario we need to assume that
the fraction of Bcd that is bound to the sites on DNA that are
specific for the hb transcription is much lower than the
concentration, $[Bcd_b]$, of Bcd bound to non-specific sites. This
occurs, in particular, if the concentration of non-specific sites is
much larger than that of sites that are specific for transcription. In
view of the hopping and sliding model of 
transcription~\cite{vonHippel15011989,Elf25052007,Hammar22062012}, this is a
reasonable assumption. We then assume that Hb and (free) Bcd are
related by (\ref{eq:Hb_hill}) with $\overline K_D \equiv
[Bcd]_{1/2}$, an effective dissociation constant between Bcd and the
sites on DNA that are specific for the transcription of hb. We show in
figure~\ref{fig:figure6}~(b) the normalized concentration of Hb as a
function of the concentrations of free and total Bcd obtained using
(\ref{eq:Hb_hill}) with $[Bcd]_{1/2}=\overline K_D=0.5\mu M$ and
$n=2$ and the distributions, $[Bcd]$ and $[Bcd_T]$, that correspond to
the stationary solution of the SDID model with partial degradation and
the parameters of table~\ref{table_1}. We note that the relationship
between $[Hb]$ and $[Bcd_T]$ in this case is more similar to the one
observed in~\cite{Gregor:Cell2007b} than the one displayed in
figure~\ref{fig:figure6}~(a). This similarity increases with increasing
values of the Hill coefficient, $n$. We here consider $n=2$ for
simplicity.  We also observe in figure~\ref{fig:figure6}~(b) that the
dependence with free Bcd is very different from the dependence with
total Bcd. Given that the observations of Bcd-GFP correspond to the
distributions of total, rather than free, Bcd, this suggests that the
Hill coefficient, $n$, that may be derived from the fluorescence
distributions may not correspond to the real cooperativity between Hb
and Bcd if a significant fraction of Bcd is bound to non-specific
sites. In order to analyze how the estimated cooperativity coefficient
may differ from the actual one if it is derived from the fluorescence
distributions under the implicit assumption that the Bcd fluorescence
is proportional to the free (not the total) Bcd concentration we
proceed as follows.  We first compute $[Hb]/[Hb]_{max}$ using
(\ref{eq:Hb_hill}) with $[Bcd]$ the free Bcd concentration and the
same parameters as in figure~\ref{fig:figure6}. We assume that this is
the relationship that holds in the real system. We then proceed as if we
had obtained the distributions of the fluorescence coming from Hb and Bcd in this system
and derive an estimated
degree of cooperativity, $n_f$,  from a fit of the form:
\begin{equation}
\frac{[Hb]}{[Hb]_{max}}=\frac{[Bcd_*]^{n_f}}{[Bcd_*]^{n_f}+[Bcd_*]_{1/2}^{n_f}},
\label{eq:Hb_hill_nf}
\end{equation}
where $[Hb]$ and $[Bcd_*]$ are proportional, respectively, to the Hb and Bcd fluorescence distributions. We compare the estimates of $n_f$ when the fluorescence is proportional to $Bcd_T$  ({\it i.e.} $Bcd_*=Bcd_T$) as we think occurs in the real system and when we use $Bcd_*=Bcd$ instead. 
To do the fitting we
rewrite (\ref{eq:Hb_hill_nf}) as:
\begin{equation}
\log\left(\frac{[Hb]_{max}}{[Hb]}-1\right)=\log\left(\frac{\overline{K}_D^{n_f}}{[Bcd_*]_{max}^{n_f}}\right) - n_f\log\left(\frac{[Bcd_*]}{[Bcd_*]_{max}}\right) ,
\label{eq:Hb_hill_log}
\end{equation}
where it becomes clear that $n_f$ is the slope of the
$\log({[Hb]_{max}}/{[Hb]}-1)$ $vs$ $\log({[Bcd_*]}{[Bcd_*]_{max}})$
relationship. We show in figure~\ref{fig:figure7}(a)
$\log({[Hb]_{max}}/{[Hb]}-1)$ as a function of
$\log({[Bcd_*]}{[Bcd_*]_{max}})$ for $[Bcd_*]$ equal to $[Bcd]$ (dashed-dotted line) or $[Bcd_T]$ (solid line). As expected, the relationship is linear in the first case
with a slope, $n_f$, that coincides with the actual cooperativity
coefficient, $n=2$.  In the other case the relationship is linear for
small $[Bcd_T]$ but, as $[Bcd_T]$ increases, the linearity is lost. In
particular, the slope changes dramatically in the region where $[Hb]$
is most sensitive to changes in $[Bcd_T]$, $[Bcd_T]/[Bcd_T]_{max} \sim
(0.5-0.8)$. If the data is fitted using (\ref{eq:Hb_hill_nf}) in
this region of great sensitivity (see figure~\ref{fig:figure7}~(a)) we
obtain $n_{f} \approx 7$ which is much larger than the actual cooperativity index,
$n=2$. The estimated Hill coefficient, $n_f$, works pretty well when
we try to reproduce the $[Hb]$ $vs$ $[Bcd_T]$ relationship over all
the range of $[Bcd_T]$ values as shown in
figure~\ref{fig:figure7}~(b). This illustrates that the inability to
distinguish between free and total Bcd can lead to an overestimation
of the Hill coefficient and of the degree of Bcd cooperativity with
which hb is transcribed.

\section{Conclusions}

Understanding the processes that lead to cell differentiation during
embryogenesis is a key goal of scientific
research~\cite{Wolpert1969,Crick1970}. Advancing in this regard is not
only relevant to improve the comprehension of how life and living
organisms are shaped but also of the limits that physics imposes on
such processes~\cite{Tkacik26082008,Dubuis08102013}. The case of the
patterning along the anterior-posterior axis of {\it Drosophila
  melanogaster} embryos is an example that has been studied in great
detail both
experimentally~\cite{Nusslein1989,Gregor:Cell2007a,Gregor:Cell2007b,Little2011,Little2013}
and through modeling~\cite{Driever1988,Bergmann2007}.  The gradient of
the protein Bicoid (Bcd) which acts as transcription factor for the
production of other proteins, is key for this process.  The Bcd
system, on the other hand, provides a paradigmatic example of the
difficulties of quantifying biophysical and biochemical parameters
from fluorescence observations.  The \textit{SDD}
model~\cite{Driever1988,Nusslein1989} was proposed to explain the
formation of the Bcd gradient in \textit{Drosophila melanogaster}
embryos but it could not account satisfactorily for all its observed
characteristics. In particular, the estimates of the Bcd diffusion
coefficient derived in~\cite{Gregor:Cell2007a} using FRAP were too
small to explain the establishment of a stable gradient within the
times observed experimentally. The estimated diffusion rate was
challenged by new measurements obtained with
FCS~\cite{AbuArish:BJ2010,Porcher2010}. These apparently contradictory
results on Bcd diffusion could be explained within a unified model
in~\cite{Sigaut2014} by considering that, in the embryo, Bcd not only
diffuses freely but also interacts with binding sites, a process that naturally
occurs in this case given that Bcd is a transcription factor. According to
this unifying model Bcd diffuses with a free coefficient, $D_f\sim
20\mu m^2/s$, and a large fraction of it is bound to immobile or very
slowly moving ($D_S\sim 0.1\mu m^2/s$) sites so that FRAP and FCS
experiments provide information on the {\it effective} coefficients of
(\ref{eq:Dsm})--(\ref{eq:Dcoll})~\cite{Sigaut2010,Sigaut2014}.  In this paper we have studied
if this type of \textit{SDID} model with the diffusion coefficients,
concentrations and dissociation constant estimated
in~\cite{Sigaut2014} can explain the formation of a Bcd gradient with
the space and time properties observed experimentally. In spite of its
simplicity, the model provides an ideal platform where to analyze how
the characteristic length and time scales of $[Bcd]$ are affected when
the interaction with binding sites is considered.  In order to
quantify some unknown parameters we compared the characteristic
lengthscale of the observed Bcd gradient with that predicted by the
model. Given that Bcd-GFP is fluorescent regardless of whether it is
free or bound we interpreted the observed lengthscale as the one that
corresponds to the total (not just the free) Bcd distribution. As
discussed in the Methods and Results sections, if Bcd binds to sites
the lengthscales of $[Bcd]$ and $[Bcd_T]$ differ by a factor $\sim
\frac{D_{sm}}{D_{coll}}$ (see (\ref{eq:z_o/z_oT})) outside the
region of Bcd synthesis. This ratio can be arbitrarily
small~\cite{Pando2006} and has been estimated to be $\sim$ 0.1 in the
anterior end of the embryo during its early
stages~\cite{Sigaut2014}. This means that the quantificaction of
biophysical parameters based on the lengthscale of the fluorescence
distribution can lead to very different parameter estimates depending
on the model with which the observations are interpreted.  Since both
$D_{sm}$ and $D_{coll}$ are nonlinear functions of the concentrations,
their ratio changes with position along the embryo. The simulations of
figure~\ref{fig:figure4} estimate it as $\sim 2$ outside the region
where Bcd is synthesized.  Although this ratio is close to one and
does not imply an order of magnitude difference in the parameter
estimates that can be derived from the observed lengthscale, according
to the simulations of figure~\ref{fig:figure4}, $[Bcd]$ and $[Bcd_T]$
decay to 50\% of their maximum values at very different distances from
the anterior end, $z \approx 100 \mu m$, and $z \approx 160 \mu m$,
respectively. This again highlights the implications that a particular
choice of model has on the interpretation of the observations.  We
based our choice of the parameters that were not estimated
in~\cite{Sigaut2014} on the fluorescence lengthscale distribution and
on the time it takes for the gradient to be established.  Although
there are some discrepancies between the results of the simulations of
figure~\ref{fig:figure4} and the experimental observations ({\it e.g.}
the lengthscale is $\sim$ 45$\mu m$ in the simulations and 100-150$\mu
m$ in the experiments) there are other processes that are not included
in figure~\ref{fig:figure4} that can bring these values closer together.
In particular, the process of GFP maturation decreases the
fluorescence intensity more pronouncedly at the anterior than at the
posterior end (see figure~\ref{fig:figure3}(b)). Thus, the
experimentally observed fluorescence lengthscale can be larger than
the actual lengthscale of the total Bcd-GFP concentration that the
simulation of figure~\ref{fig:figure4} represents.

The simplicity of the SDD model is very appealing. Within its context,
the properties that are observed experimentally are directly related
to the parameters of the model.  The SDID model is nonlinear and a
direct comparison between theory and experiment is more
complicated. In spite of this, in this paper we went beyond the
numerical simulations and obtained analytical expressions that could
describe the simulated results. In this way we could establish that
the role that the free diffusion coefficient plays in the SDD model,
in the SDID model it is played by the largest of the two effective
diffusion coefficients of~\cite{Pando2006} (the collective coefficient
of (\ref{eq:Dcoll})) as illustrated in
figure~\ref{fig:figure4}--\ref{fig:figure5}. Had it been the single
molecule coefficient that is estimated with FRAP~\cite{Sigaut2010},
the timescale of the gradient formation would have been too large
compared to the experimental observations. This is especially
important for the action of Bcd as transcription factor and the
precision with which its ``bulk'' concentration can be estimated by
the regulatory binding sites on
DNA~\cite{Gregor:Cell2007b,Ipina2016}. Considering the interaction of
Bcd with binding sites as done in our SDID model has major
consequences for the interpretation of the experiments that seek to
quantify the action of Bcd as transcription factor. More specifically,
given that the fluorescence does not distinguish between free and
bound Bcd, the relationship between the Bcd concentration and that of
the proteins, {\it e.g.}  Hunchback (Hb), whose production it
regulates needs to be reassessed. As shown in figure~\ref{fig:figure6}
the SDID model predicts that the Hill coefficient that characterizes
the cooperativity with which Bcd promotes the production of Hb can be
smaller than the one that is directly derived from the scatter plot of
the observed fluorescence. This conclusion is derived under the
assumption that the fraction of regulatory sites on DNA that Bcd binds
to is negligible compared to those that are included in the SDID model
which should then correspond to other (non-specific) binding sites. In
view of the hopping and sliding model of
transcription~\cite{vonHippel15011989} and the typical fraction of
time that transcription factors spend bound to non-specific
sites~\cite{Elf25052007,Hammar22062012} it is very likely that this
assumption be valid in the case of Bcd. Given that most intracellular
messengers are likely to be subject to binding/unbinding processes, it
is likely that similar problems to those discussed here will be found in
other systems as well.  Our results are then not only relevant for the
particular case of the gradient of Bcd but also have wide implications
for the interpretation of fluorescence images in living organisms in
general.

\section*{Acknowledgments}
This research has been supported by Universidad de Buenos Aires
(UBACyT 20020130100480BA), Agencia Nacional de Promoci\'on
Cient\'ifica y Tecnol\'ogica (PICT 2013-1301).  SPD
is a member of Carrera del Investigador Cient\'ifico (Consejo Nacional
de Investigaciones Cient\'ificas y T\'ecnicas).

\section*{Appendix}
In this Appendix we present the  versions of the SDID model that we 
have implemented and analyzed but have not described in detail in the manuscript. 

\subsection*{The SDID model with total degradation of Bcd.} 
In this case we consider that Bcd is degraded while being free or bound and the dynamic equations read:
\begin{equation}
\eqalign{
\frac{\partial{[Bcd]}}{\partial t} = & D_{f}\nabla^2 [Bcd] -k_{on}[Bcd]([S_T] - [Bcd_b]) +  k_{off}[Bcd_b] \\
&-\overline{\alpha}[Bcd] + \theta(z),\\ 
\frac{\partial{[Bcd_b]}}{\partial t}  = & D_{S}\nabla^2 [Bcd_b] +k_{on}[Bcd]([S_T] - [Bcd_b]) - k_{off}[Bcd_b] \\ & -\overline{\alpha} [Bcd_b] ,}
\label{eq:rd_eq_deg_total}
\end{equation}
where $\overline{\alpha}$ is the degradation rate.  The choice of parameter
values is done as in the model with partial degradation.  We
first derive the stationary solution for $D_S=0$. In this case the
equilibrium condition (\ref{eq:eq_Bcdb})--(\ref{eq:eq_S}) does not
hold for all $z\ge 0.2L$. But if reactions occur on a faster
time-scale than degradation (as in the ``fast reaction approximation''
of (\ref{eq:Bcd_fast_buffers_Ds_0})), it is possible to assume
that (\ref{eq:eq_Bcdb})--(\ref{eq:eq_S}) hold approximately at
every $z\in [0.2L, L]$. In such a case, the evolution equation for
$[Bcd]$ is given by (\ref{eq:Bcd_fast_buffers_Ds_0}) with 
%\begin{eqnarray}
%\frac{\partial{[Bcd]}}{\partial t} & = & D_{coll}\nabla^2 [Bcd] -\tilde{\alpha}[Bcd] + \hat{\theta}(z),
%\label{eq:Bcd_fast_buffers_Ds_0_two_alpha}
%\end{eqnarray}
 $\tilde{\alpha}=\overline{\alpha} D_{coll}/D_{sm}$ and  $\hat{\theta}$ as before.  As in
the model with partial degradation
we estimate the lengthscales of the stationary solution as:
\begin{equation}
z_{o_f}'\approx \sqrt{D_{coll}/\tilde{\alpha}}=\sqrt{D_{sm}/\overline{\alpha}},
\label{eq:zof_deg_tot}
\end{equation}
for $[Bcd]$, and,
\begin{equation}
z_{o_T}'=z_{o_f}'D_{coll}/D_{sm},
\label{eq:z_o/z_oT_deg_tot}
\end{equation} 
for $[Bcd_T]$. Although the ratio between the characteristic
lengthscales of $[Bcd]$ and $[Bcd_T]$ in this case is given by
(\ref{eq:z_o/z_oT}) as in the partial degradation model, the
lengthscale of the gradient depends on different biophysical
parameters. The estimate of the degradation rate, 
 $\overline{\alpha}$, that may be derived  from the characteristic lengthscale
of $[Bcd_T]$ in this case is approximately related
to the one obtained in the model with partial degradation by $\overline{\alpha}=\tilde{\alpha}
D_{sm}/D_f$. Since $D_{sm}\ll D_f$, if we use this value of
$\overline{\alpha}$ and the model with total Bcd
degradation to determine the source intensity as before we obtain a
value, $\theta$, that is smaller by a factor, $D_{sm}/D_f$, with
respect to the one derived using the partial degradation model.
Taking into account that in the region where FCS experiments are
performed $D_{sm}/D_f\sim 0.05$, using
(\ref{eq:zof_deg_tot})--(\ref{eq:z_o/z_oT_deg_tot}) we obtain
the \textit{a priori} estimate $\overline{\alpha} \sim
0.005s^{-1}$. The numerical simulations performed with this value did
not give proper concentration distributions for the different
species. Hence we used $\overline{\alpha}=0.0005 s^{-1}$ and
$\overline{\theta}_0=0.1 nM s^{-1}$ instead.

\subsection*{The SDID model with with GFP maturation}
Experiments use Bcd-GFP to observe the distribution of Bcd. It takes
some time for GFP to mature and become
fluorescent~\cite{Sniegowski2005,Iizuka2011,Little2011}. Thus, to
interpret the observations it may be necessary to include this
process. In such a case we need to distinguish between fluorescent (or
{\it tagged}) and non-fluorescent (or {\it untagged}) Bcd ($Bcd^{t}$
and $Bcd^{u}$, respectively) and include the transformation between
one another. The equations then read:
%\begin{eqnarray}
%\frac{\partial{[Bcd^{u}]}}{\partial t}& = & D_{f}\nabla^2 [Bcd^{u}] -k_{on}[Bcd^{u}]\left([S_T] - [Bcd_b^{u}]- [Bcd_b^{t}]\right) + 
%k_{off}[Bcd_b^{u}] -\alpha[Bcd^{u}] - \gamma [Bcd^{u}],\\ 
%\frac{\partial{[Bcd^{t}]}}{\partial t}& = & D_{f}\nabla^2 [Bcd^{t}] -k_{on}[Bcd^{t}]\left([S_T] - [Bcd_b^{u}]- [Bcd_b^{t}]\right) + 
%k_{off}[Bcd_b^{t}] -\alpha[Bcd^{t}] + \gamma [Bcd^{u}],\\ 
%\frac{\partial{[Bcd_b^{u}]}}{\partial t}& = & D_{S}\nabla^2 [Bcd_b^{u}] +k_{on}[Bcd^{u}]\left([S_T] - [Bcd_b^{u}]- [Bcd_b^{t}]\right)-k_{off}[Bcd_b^{u}]- \gamma [Bcd_b^{u}],\\
%\frac{\partial{[Bcd_b^{t}]}}{\partial t}& = & D_{S}\nabla^2 [Bcd_b^{t}] +k_{on}[Bcd^{t}]\left([S_T] - [Bcd_b^{u}]- [Bcd_b^{t}]\right)-k_{off}[Bcd_b^{t}]+\gamma [Bcd_b^{u}],
%\label{eq:rd_eq_tagged_untagged}
%\end{eqnarray}
\begin{equation}
\eqalign{
\frac{\partial{[Bcd^{u}]}}{\partial t} = & D_{f}\nabla^2 [Bcd^{u}] -k_{on}[Bcd^{u}]\left([S_T] - [Bcd_b^{u}]- [Bcd_b^{t}]\right) \\ & + k_{off}[Bcd_b^{u}] -\tilde{\alpha}[Bcd^{u}] - \gamma [Bcd^{u}] + \theta(z),\\ 
\frac{\partial{[Bcd^{t}]}}{\partial t} = & D_{f}\nabla^2 [Bcd^{t}] -k_{on}[Bcd^{t}]\left([S_T] - [Bcd_b^{u}]- [Bcd_b^{t}]\right) \\ & + k_{off}[Bcd_b^{t}] -\tilde{\alpha}[Bcd^{t}] + \gamma [Bcd^{u}],\\ 
\frac{\partial{[Bcd_b^{u}]}}{\partial t} = & D_{S}\nabla^2 [Bcd_b^{u}] +k_{on}[Bcd^{u}]\left([S_T] - [Bcd_b^{u}]- [Bcd_b^{t}]\right) \\ &-k_{off}[Bcd_b^{u}] - \gamma [Bcd_b^{u}],\\
\frac{\partial{[Bcd_b^{t}]}}{\partial t} = & D_{S}\nabla^2 [Bcd_b^{t}] +k_{on}[Bcd^{t}]\left([S_T] - [Bcd_b^{u}]- [Bcd_b^{t}]\right) \\ & -k_{off}[Bcd_b^{t}]  + \gamma [Bcd_b^{u}],
}
\label{eq:rd_eq_tagged_untagged}
\end{equation}
where $\gamma$ is the rate of GFP maturation. Here we assume that this
maturation only affects whether the protein is fluorescent or not but
not the properties of its transport, binding or degradation. We also
assume that immediately after its synthesis the protein is not
fluorescent.

% \textcolor{red}{ACA
                                                                                                                                                                                                                                                                                              % PONER
                                                                                                                                                                                                                                                                                              % LO
                                                                                                                                                                                                                                                                                              % DE
                                                                                                                                                                                                                                                                                              % liu
                                                                                                                                                                                                                                                                                              % et
                                                                                                                                                                                                                                                                                              % al?}.

\section*{References}
\bibliography{bcd_gradient_2}

\begin{thebibliography}{}

\bibitem[Abu-Arish et~al., 2010]{AbuArish:BJ2010}
Abu-Arish, A., Porcher, A., Czerwonka, A., Dostatni, N., and Fradin, C. (2010).
\newblock High mobility of bicoid captured by fluorescence correlation
  spectroscopy: Implication for the rapid establishment of its gradient.
\newblock {\em Biophysical Journal}, 99(4):L33--L35.

\bibitem[Bergmann et~al., 2007]{Bergmann2007}
Bergmann, S., Sandler, O., Sberro, H., Shnider, S., Schejter, E., Shilo, B.-Z.,
  and Barkai, N. (2007).
\newblock Pre-steady-state decoding of the bicoid morphogen gradient.
\newblock {\em PLoS Biol}, 5(2):e46.

\bibitem[Crick, 1970]{Crick1970}
Crick, F. (1970).
\newblock Diffusion in {Etibryogenesis}.
\newblock {\em Nature}, 225:421.

\bibitem[Douglas and Rachford, 1956]{Douglas-Rachford_ADI}
Douglas, J. and Rachford, H.~H. (1956).
\newblock On the numerical solution of heat conduction problems in two and
  three space variables.
\newblock {\em Transactions of the American mathematical Society}, pages
  421--439.

\bibitem[Driever and Nusslein-Volhard, 1989]{Nusslein1989}
Driever, W. and Nusslein-Volhard, C. (1989).
\newblock The bicoid protein is a positive regulator of hunchback transcription
  in the early drosophila embryo.
\newblock {\em Nature}, 337:138--143.

\bibitem[Driever and Nussleinvolhard, 1988]{Driever1988}
Driever, W. and Nussleinvolhard, C. (1988).
\newblock Stability and nuclear dynamics of the bicoid morphogen gradient.
\newblock {\em Cell}, 54(1):83--93.

\bibitem[Drocco et~al., 2011]{Drocco2011}
Drocco, J.~A., Grimm, O., Tank, D.~W., and Wieschaus, E. (2011).
\newblock Measurement and perturbation of morphogen lifetime: effects on
  gradient shape.
\newblock {\em Biophysical journal}, 101(8):1807--1815.

\bibitem[Dubuis et~al., 2013]{Dubuis08102013}
Dubuis, J.~O., Tkačik, G., Wieschaus, E.~F., Gregor, T., and Bialek, W.
  (2013).
\newblock Positional information, in bits.
\newblock {\em Proceedings of the National Academy of Sciences},
  110(41):16301--16308.

\bibitem[Elf et~al., 2007]{Elf25052007}
Elf, J., Li, G.-W., and Xie, X.~S. (2007).
\newblock Probing transcription factor dynamics at the single-molecule level in
  a living cell.
\newblock {\em Science}, 316(5828):1191--1194.

\bibitem[Gregor et~al., 2007a]{Gregor:Cell2007a}
Gregor, T., Wieschaus, E.~F., McGregor, A.~P., Bialek, W., and Tank, D.~W.
  (2007a).
\newblock Stability and nuclear dynamics of the bicoid morphogen gradient.
\newblock {\em Cell}, 130(1):141--152.

\bibitem[Gregor et~al., 2007b]{Gregor:Cell2007b}
Gregor, T., T., D.W., W., and E.F., Bialek, W. (2007b).
\newblock Probing the limits to positional information.
\newblock {\em Cell}, 130(1):153--164.

\bibitem[Grimm et~al., 2010]{Grimm2010}
Grimm, O., Coppey, M., and Wieschaus, E. (2010).
\newblock Modelling the bicoid gradient.
\newblock {\em Development}, 137(14):2253--2264.

\bibitem[Hammar et~al., 2012]{Hammar22062012}
Hammar, P., Leroy, P., Mahmutovic, A., Marklund, E.~G., Berg, O.~G., and Elf,
  J. (2012).
\newblock The lac repressor displays facilitated diffusion in living cells.
\newblock {\em Science}, 336(6088):1595--1598.

\bibitem[Hecht et~al., 2009]{Hecht2009}
Hecht, I., Rappel, W.-J., and Levine, H. (2009).
\newblock Determining the scale of the bicoid morphogen gradient.
\newblock {\em Proceedings of the National Academy of Sciences},
  106(6):1710--1715.

\bibitem[Houchmandzadeh et~al., 2002]{Houchmandzadeh2002}
Houchmandzadeh, B., Wieschaus, E., and Leibler, S. (2002).
\newblock Establishment of developmental precision and proportions in the early
  {Drosophila} embryo.
\newblock {\em Nature}, 415(6873):798--802.

\bibitem[Iizuka et~al., 2011]{Iizuka2011}
Iizuka, R., Yamagishi-Shirasaki, M., and Funatsu, T. (2011).
\newblock Kinetic study of de novo chromophore maturation of fluorescent
  proteins.
\newblock {\em Analytical biochemistry}, 414(2):173--178.

\bibitem[Ipi\~na and Dawson, 2016]{Ipina2016}
Ipi\~na, E.~P. and Dawson, S.~P. (2016).
\newblock Fluctuations, correlations and the estimation of concentrations
  inside cells.
\newblock {\em PLoS ONE}, 11(3):e0151132.

\bibitem[Little et~al., 2013]{Little2013}
Little, S.~C., Tikhonov, M., and Gregor, T. (2013).
\newblock Precise developmental gene expression arises from globally stochastic
  transcriptional activity.
\newblock {\em Cell}, 154(4):789--800.

\bibitem[Little et~al., 2011]{Little2011}
Little, S.~C., Tka{\v{c}}ik, G., Kneeland, T.~B., Wieschaus, E.~F., and Gregor,
  T. (2011).
\newblock The formation of the bicoid morphogen gradient requires protein
  movement from anteriorly localized mrna.
\newblock {\em PLoS biology}, 9(3):e1000596.

\bibitem[Liu et~al., 2013]{Liu2013}
Liu, F., Morrison, A.~H., and Gregor, T. (2013).
\newblock Dynamic interpretation of maternal inputs by the drosophila
  segmentation gene network.
\newblock {\em Proceedings of the National Academy of Sciences},
  110(17):6724--6729.

\bibitem[Milo and Phillips, 2015]{Bionumbers}
Milo, R. and Phillips, R. (2015).
\newblock {\em Cell Biology by the numbers - DRAFT}.
\newblock Garland Science.

\bibitem[Pando et~al., 2006]{Pando2006}
Pando, B., Dawson, S.~P., Mak, D.-O.~D., and Pearson, J.~E. (2006).
\newblock Messages diffuse faster than messengers.
\newblock {\em Proc Natl Acad Sci (USA)}, 103(14):5338--5342.

\bibitem[Porcher et~al., 2010]{Porcher2010}
Porcher, A., Abu-Arish, A., Huart, S., Roelens, B., Fradin, C., and Dostatni,
  N. (2010).
\newblock The time to measure positional information: maternal hunchback is
  required for the synchrony of the bicoid transcriptional response at the
  onset of zygotic transcription.
\newblock {\em Development}, 137(16):2795--2804.

\bibitem[Sigaut et~al., 2014]{Sigaut2014}
Sigaut, L., Pearson, J.~E., Colman-Lerner, A., and Ponce~Dawson, S. (2014).
\newblock Messages do diffuse faster than messengers: Reconciling disparate
  estimates of the morphogen bicoid diffusion coefficient.
\newblock {\em PLoS Comput Biol}, 10(6):e1003629.

\bibitem[Sigaut et~al., 2010]{Sigaut2010}
Sigaut, L., Ponce, M.~L., Colman-Lerner, A., and Dawson, S.~P. (2010).
\newblock Optical techniques provide information on various effective diffusion
  coefficients in the presence of traps.
\newblock {\em Phys. Rev. E}, 82:051912.

\bibitem[Sniegowski et~al., 2005]{Sniegowski2005}
Sniegowski, J.~A., Phail, M.~E., and Wachter, R.~M. (2005).
\newblock Maturation efficiency, trypsin sensitivity, and optical properties of
  arg96, glu222, and gly67 variants of green fluorescent protein.
\newblock {\em Biochemical and biophysical research communications},
  332(3):657--663.

\bibitem[Spirov et~al., 2009]{Spirov2009}
Spirov, A., Fahmy, K., Schneider, M., Frei, E., Noll, M., and Baumgartner, S.
  (2009).
\newblock Formation of the bicoid morphogen gradient: an {mRNA} gradient
  dictates the protein gradient.
\newblock {\em Development}, 136(4):605--614.

\bibitem[Sprague and McNally, 2005]{Sprague2005}
Sprague, B. and McNally, J. (2005).
\newblock Frap analysis of binding: Proper and fitting.
\newblock {\em Trends in Cell Biology}, 15(2):84--91.

\bibitem[St~Johnston et~al., 1989]{st_johnston1989}
St~Johnston, D., Driever, W., Berleth, T., Richstein, S., and
  Nüsslein-Volhard, C. (1989).
\newblock Multiple steps in the localization of bicoid {RNA} to the anterior
  pole of the {Drosophila} oocyte.
\newblock {\em Development (Cambridge, England)}, 107 Suppl:13--19.

\bibitem[Strier and Dawson, 2000]{strier2000}
Strier, D.~E. and Dawson, S.~P. (2000).
\newblock Rescaling of diffusion coefficients in two-time scale chemical
  systems.
\newblock {\em The Journal of Chemical Physics}, 112(2):825--834.

\bibitem[Struhl et~al., 1989]{Strhulet1989}
Struhl, G., Struhl, K., and Macdonald, P.~M. (1989).
\newblock The gradient morphogen bicoid is a concentration-dependent
  transcriptional activator.
\newblock {\em Cell}, 57:1259--1273.

\bibitem[Tkačik et~al., 2008]{Tkacik26082008}
Tkačik, G., Callan, C.~G., and Bialek, W. (2008).
\newblock Information flow and optimization in transcriptional regulation.
\newblock {\em Proceedings of the National Academy of Sciences},
  105(34):12265--12270.

\bibitem[von Hippel and Berg, 1989]{vonHippel15011989}
von Hippel, P.~H. and Berg, O.~G. (1989).
\newblock Facilitated target location in biological systems.
\newblock {\em Journal of Biological Chemistry}, 264(2):675--678.

\bibitem[Wagner and Keizer, 1994]{Wagner:Biophysj1994}
Wagner, J. and Keizer, J. (1994).
\newblock Effects of rapid buffers on ca2+ diffusion and ca2+ oscillations.
\newblock {\em Biophys. J.}, 67(1):447--456.

\bibitem[White et~al., 2016]{White201675}
White, M., Angiolini, J., Alvarez, Y., Kaur, G., Zhao, Z., Mocskos, E., Bruno,
  L., Bissiere, S., Levi, V., and Plachta, N. (2016).
\newblock Long-lived binding of sox2 to \{DNA\} predicts cell fate in the
  four-cell mouse embryo.
\newblock {\em Cell}, 165(1):75 -- 87.

\bibitem[Wolpert, 1969]{Wolpert1969}
Wolpert, L. (1969).
\newblock Positional information and the spatial pattern of cellular
  differentiation.
\newblock {\em Journal of Theoretical Biology}, 25(1):1--47.

\end{thebibliography}
%\bibliographystyle{bcd_gradient}
%\endbib
\end{document}